\documentclass[onecolumn,12pt]{els-mrw} 
\usepackage{amsmath,amssymb,amsfonts,amsthm,makeidx,graphicx}
\usepackage{txfonts}
\usepackage{helvet}
\newcommand{\be}{\begin{equation}}
\newcommand{\ee}{\end{equation}}
\newcommand{\tr}{{\rm Tr}}

\begin{document}

\chapter{Semiclassical theory of transport}\label{chap1}

\author[1]{Marcel Novaes}%
\address[1]{\orgname{Universidade Federal de Uberl\^andia}, \orgdiv{Instituto de F\'isica}, \orgaddress{38408-100, Uberl\^andia, MG, Brazil}}

\articletag{Chapter Article tagline: Sept. 1, 2025}

\maketitle

\begin{glossary}[Keywords]
Quantum Chaos, Semiclassical Approximation, Transport, Time Delay

\end{glossary}

\begin{abstract}[Abstract]
We discuss the semiclassical approximation to transport problems in quantum chaotic systems. The figures of merit are moments of the transmission matrix and of the time delay matrix. After reviewing a few results obtained by treating these matrices are random matrices, we show how expressions for their  elements in terms of sums over trajectories lead to diagrammatic formulations that correspond to perturbative calculations. This semiclassical approach agrees with random matrix theory when it should, and allows further elements to be incorporated, like tunnel barriers, superconductors, absorption effects. We also discuss how this approach can be encoded in matrix integrals, resulting in a powerful  and versatile theory that is amenable to algebraic solutions. 
\end{abstract}

\section{Introduction}\label{chap1:sec1}

The scattering problem we have in mind starts with a two-dimensional billiard, a certain region in the plane inside of which movement is ballistic and reflections off the wall are specular, leading to chaotic dynamics. We then attach to this region two infinite waveguides, called leads. We consider only two leads for simplicity, but everything can be generalized to a larger number. The cross-section of each lead is assumed constant, so that the Schrödinger equation is separable within them. In the longitudinal direction we have free propagation, while in the transverse direction we have quantization, leading to the eigenstates of a one-dimensional particle in a box. If the total energy $E$ is fixed, then a finite number of transversal modes can be populated in each lead, say $N_1$ and $N_2$ (these numbers scale as $\sqrt{E}w/\hbar$, where $w$ is the lead width). 

Scattering of waves by the chaotic region is described by a matrix $S$ that connects the incoming wave function amplitudes in the transverse channels with the outgoing amplitudes,
\be\label{S} \begin{pmatrix}b_1\\b_2\end{pmatrix} =S\begin{pmatrix}a_1\\a_2\end{pmatrix}.\ee
In this equation $a_i$ and $b_i$ are vectors of dimension $N_i$, while $S$ has four blocks
\be S=\begin{pmatrix}r&t'\\t&r'\end{pmatrix}.\ee
Without loss of generality, we assume that $N_2\ge N_1$. We have in mind non-interacting electrons.

The block $r$ is $N_1\times N_1$ and measures how much of the waves incident on lead $1$ are reflected back; on the other hand, the block $t$ is $N_2\times N_1$ and measures how much of the waves incident from lead $1$ are transmitted through lead 2; the role of the other blocks is analogous. Probability conservation is guaranteed by the fact that the matrix $S$ is unitary,
\be S^\dagger S =1.\ee

The transport process can also be studied in terms of the Hermitian transmission matrix
\be\label{T} T=t^\dagger t.\ee 
Its $N_1$ real eigenvalues are between $0$ and $1$ and can be interpreted as transmission probabilities. Analogously, the eigenvalues of $r^\dagger r$ can be interpreted as reflection probabilities. These eigenvalues are subject to the constraint $t^\dagger t+r^\dagger r=1$.

The total transmission is given by the trace ${\rm Tr}(T)$ and is called the (dimensionless) conductance of the system. If a small voltage is applied between the leads, a time-dependent electric current will flow, $I(t)$. This typically fluctuates widely as a function of the time, and the conductance corresponds to its average over a long time window. Further information about transport can be obtained from the other traces, ${\rm Tr}(T^n)$. For example, even at zero temperature there is noise in quantum transport, due to the granular nature of the electric charge (each electron has a probability of being transmitted and a probability of being reflected, generating quantum oscillations in the current). This noise, called shot noise, corresponds to the variance of current and is measured by ${\rm Tr}(T(1-T))$. Traces of matrix polynomials in $T$ are called transport moments.

In the 1950's, Eugene Wigner introduced a random matrix model for the study of nuclear physics. Instead of trying to solve the problem exactly, or numerically, in order to obtain precise energy levels of a nucleus, he argued that the Hamiltonian of this many-body system would be so complicated, so sensitive to variation of parameters over which there is no control, that it would make sense to simply replace it with a random matrix. Wigner thus inaugurated a statistical approach to nuclear physics in which the exact values of quantities are less important than their distribution. \footnote{Refer to chapter X in this volume}

In the 1980s, some studies suggested that this approach could be used in quantum chaos, since the same unpredictability argument used for many-body systems would also apply to single-body systems, provided that they had chaotic dynamics and were operating in a regime where the classical dynamics was relevant, i.e. the wavelength was much smaller than the system size, $\lambda\ll L$. This idea later became known as the Bohigas, Giannoni, and Schmit conjecture. We then consider an ensemble of systems, all of them chaotic but with the same macroscopic specifications such as area or numbers of channels in the leads, and than calculate the average value of observables over this ensemble. It is like classical statistical mechanics, but instead of replacing position and momentum by random numbers, we replace observables by random operators, represented by random matrices. Results are then expected to be independent of specific characteristics of the system, instead being determined only by some macroscopic parameters like $N_1, N_2$, whether the dynamics is invariant under time-reversal, and whether the electronic spin is relevant. This phenomenon is known as universality \cite{Jalabert1990,Blumel1990}.

The type of random matrix to be considered depends on the constraints that need to be respected. If it is the Hamiltonian of a closed system, it must be Hermitian and, in the presence of time-reversal, it must be real; the $S$ matrix of a scattering problem must be unitary and, in the presence of time-reversal, it must be symmetric \cite{Beenakker1997}; the transmission matrix $T$ is hermitian with eigenvalues between $0$ and $1$.

On the other hand, a semiclassical approximation can also be justified when $\lambda\ll L$, when waves can be treated like geometrical rays \cite{Berry1985}. In this regime, it is reasonable to expect that calculations that are based on classical quantities might lead to good approximate results, at least in some regimes. In closed systems, universality requires that the Heisenberg time $\tau_H$, defined as Planck's constant divided by the mean level spacing, be very large, in particular compared to the period of the shortest periodic orbit and the inverse Lyapunov exponent, $\lambda_L^{-1}$. In open systems, other time scales are present, like the dwell time, $\tau_D$, defined as the average time spent withing the scattering region by a particle injected at random. The physical hypothesis underlying universality is that, once a particle enters the cavity, its wavefunction will spread all around, leading to reflection and transmission amplitudes that are essentially random and do not depend on details of how the particle started its motion. This requires that $\tau_D$ be large enough, certainly $\tau_D\gg \lambda_L^{-1}$. This means that the leads must be small compared to the system size.

On the other hand, in the semiclassical limit a wavepacket stays localized for longer times. Physically, if the wavelength is much smaller than the openings, noiseless transmission becomes possible, i.e., states may enter and leave the cavity before they are efficiently randomized. This effect is measured by the so-called Ehrenfest time, $\tau_E=\lambda_L^{-1}\log(c/\hbar)$, where $c$ is related to the size of the openings and its precise value depends on the specific situation being considered \cite{Schomerus_2005,PhysRevB.73.195115}. This can lead to a suppression of shot-noise, for instance, that scales like $e^{-\tau_E/\tau_D}$.  

The universality regime is then $\tau_H\gg\tau_D$, $\tau_D\gg \tau_E$, $\tau_D\gg \lambda_L^{-1}$, and in this regime we should expect random matrix theory and the semiclassical approach to be effective and give the same results. However, transport moments for chaotic systems are in practice strongly fluctating as functions of energy, so on the semiclassical side local energy averages must be carried out in order to arrive at really universal results (this local energy average is not unlike the ensemble average that underlies the RMT side). 

Of course all of the above assumes a completely chaotic dynamics. When a system is more generic, with regions of regular and chaotic motion coexisting in phase space, its dynamics will be much more rich and intricate and this impacts its quantum properties \cite{pr}. A fractal hierarchy of stability islands may exist, as well as fractal invariant sets called cantori. Chaotic trajectories that come close to these structures may experience the phenomenon of stickiness, i.e. become trapped for a duration of time that varies erratically with the initial condition. Quantum mechanically this gives rise to resonance-assisted tunneling \cite{rat,rat2} or reduced quantum transport due to partial barriers \cite{partial1, partial2}.

A different characterization of quantum chaotic scattering is in terms of the time delay operator, defined as $Q=-i\hbar S^{\dagger}\frac{dS}{dE}$, in terms of the energy derivative of the $S$ matrix. In this context it makes sense to consider all \be M=N_1+N_2\ee open channels at once and disregard the block structure of $S$. The normalized trace (average eigenvalue) $\tau_W=M^{-1}{\rm Tr}(Q)$ is called the Wigner time delay. Its local energy average equals the classical dwell time, $\langle \tau_W\rangle=\tau_D$. More detailed information about the time spent by the particle in the scattering region can be obtained from other traces ${\rm Tr}(Q^n)$. Again, the problem may be addressed using randon matrix theory \cite{Lehmann1995,fyodorov,sokolov,texier} and also the semiclassical approximation \cite{Eckhardt,Raul,Caio,PhysRevE.77.046219}.

This chapter is organized as follows. In Section II we outline the RMT approach to quantum transport (a topic that is covered in much more detail in another chapter). In Section III we discuss the semiclassical approach, its diagrammatic formulation and how it lead to the first results showing equivalence with RMT in the universality regime. Section IV is dedicated to an efficient formulation of the semiclassical approach in terms of matrix integrals, which leads to a much deeper understanding of said equivalence.

\section{Random matrix theory}

For the scattering problem, the statistical approach prescribes replacing the $S$ matrix by a random unitary matrix. This can be done using the uniform probability distribution that is naturally defined over the unitary group, known as the Haar measure. This ensemble is called CUE, Circular Unitary Ensemble. In the presence of time reversal symmetry, the matrix $S$ must be symmetric and is taken with uniform distribution over the COE, Circular Orthogonal Ensemble. We shall restrict our attention, for simplicity, to the case of broken time reversal symmetry. \footnote{Again, refer to chapter X in this volume}

When $S$ is uniformly distributed over the CUE, this induces a probability distribution for the transmission matrix $T$ defined in (\ref{T}). It turns out that this distribution is a particular case of the Jacobi ensemble \cite{jacobi}. The joint distribution of eigenvalues, in this case, is given by
\be\label{dt} P(T)\propto |\Delta(T)|^\beta \prod_{i=1}^{N_1}T_i^{\frac{\beta}{2}(N_2-N_1+1)-1},\ee
where 
\be \Delta(T)=\prod_{i<j}(T_j-T_i)\ee
is called the Vandermonde of variables $T$ and the Dyson index $\beta$ is equal to $1$, $2$ or $4$ depending on whether the matrix elements are real, complex or quaternions. The normalization of this distribution results from the integral
\be \int_0^1 dT \Delta(T)^{2c} \prod_{i=1}^{N_1}T_i^{a-1}(1-T_i)^{b-1}=\prod_{j=0}^{N_1-1}\frac{\Gamma(1+c+jc)\Gamma(a+jc)\Gamma(b+jc)}{\Gamma(1+c)\Gamma(a+b+(N_1+j-1)c)},\ee known as the Selberg integral \cite{Forrester2008}. The presence of the Vandermonde shows that the eigenvalues of $T$ are subject to so-called level repulsion. 

With the joint distribution of eigenvalues (\ref{dt}) we can calculate all kinds of statistical characteristics of transport. The average conductance of the system, for example, is $\langle {\rm Tr}(T)\rangle=\frac{N_1N_2}{M+2/\beta-1}$, $M$ being the total number of channels. The conductance variance, in turn, which describes the phenomenon of universal conductance fluctuations, is given when $\beta=2$ by $\langle ({\rm Tr}(T))^2\rangle-\langle {\rm Tr}(T)\rangle^2=\frac{N_1^2N_2^2}{M^2(M^2-1)}.$ Early results obtained using the description via random matrices, mostly valid either for very small or very large channel numbers, were collected in \cite{Beenakker1997}.

From now on we focus for simplicity on the case $\beta=2$.

An efficient approach to computing these transport quantities was introduced in \cite{Novaes2008} by relating them to a very important set of polynomials, the Schur polynomials $s_\lambda(T)$, computed at the eigenvalues of $T$. They are identified by partitions, non-increasing sequences of positive integers $\lambda=(\lambda_1,\lambda_2,...)$, and are a fundamental ingredient of the representation theory of both permutation groups and unitary groups \cite{mcdonald}. The mean value of a Schur polynomial of the transmission eigenvalues is given by a Selberg-like integral \cite{kaneko,kadell},
\be \langle s_\lambda(T)\rangle=\frac{s_\lambda(1_{N_1})s_\lambda(1_{N_2})}{n!s_\lambda(1_{M})}.\ee 

In the above expression use is made of the value of the Schur polynomial evaluated at identity matrices of different dimensions. There is an explicit formula for this quantity. Let $(N)^{(n)}=N(N+1)\cdots (N+n-1)$ be the raising factorial, and let $[M]^{(\lambda)}=\prod_{i=1}^{\ell(\lambda)}(M-i+1)^{(\lambda_i)}$ be a generalization of it. Let $\sum_i \lambda_i=n$ and $\ell$ be the number of parts in $\lambda$. Then
\be s_\lambda(1_{M})=\frac{d_\lambda}{n!}[M]^{(\lambda)},\ee
where
\be d_\lambda=|\lambda|!\prod_{i=1}^{\ell}\frac{1}{(\lambda_i-i+\ell)!}\prod_{j=i+1}^{\ell}(\lambda_i-\lambda_j+j-i)\ee is the dimension of the irreducible representation of the permutation group which corresponds to the partition $\lambda$ (the quantity $s_\lambda(1_{M})$ is also the dimension of an irreducible representation, but of the unitary group).

With this, it is possible to calculate observables like the moments of conductance or of shot noise \cite{Savin2006}. For example, 
\be \left\langle{\rm Tr}(T^n)\right\rangle=\sum_{p=0}^{n-1}\frac{(-1)^p}{n!}\binom{n-1}{p}\frac{(N_1-p)^{(n)}(N_2-p)^{(n)}}{(M-p)^{(n)}}.\ee

On the other hand, eigenvalues of the time delay operator are unbounded positive, and the adequate random matrix ensemble for them is the inverse Laguerre \cite{Brouwer1997}, such that
\be P(Q)\propto \frac{|\Delta(Q)|^2}{\det(Q)^{3M-2}}e^{-M\tau_D{\rm Tr}(Q^{-1})}.\ee Upon a change of variable, its normalization follows from the integral
\be \int |\Delta(\Gamma)|^2\det(\Gamma)^Me^{-M\tau_D{\rm Tr}(\Gamma)}=\frac{1}{(M\tau_D)^{2M^2}}\prod_{i=1}^M i!(M+i-1)!.\ee

Linear statistics can be derived from the density of eigenvalues of $Q$ \cite{savin1,savin2,savin3,cunden}, and more general moments are related to Selberg-like integrals \cite{mezz,tau,Cunden2016,Cunden2016b}. Introducing again Schur polynomials, and resorting to another generalization of the Selberg integral \cite{Novaes2015a}, one can show that 
\be\label{schurQ} \langle s_\lambda(Q)\rangle=(M\tau_D)^{|\lambda|}\frac{d_\lambda}{|\lambda|!}\frac{[M]^\lambda}{[M]_\lambda},\ee
where $ [M]_{(\lambda)}=\prod_{i=1}^{\ell(\lambda)}(M+i-1)_{(\lambda_i)}$
is a generalization of the falling factorial. For example, 
\be \left\langle{\rm Tr}(Q^n)\right\rangle=\frac{(M\tau_D)^n}{n!}\sum_{p=0}^{n-1}\frac{(-1)^p}{n!}\binom{n-1}{p}\frac{(M-p)^{(n)}}{(M+p)_{(n)}}.\ee

It is interesting how the random matrix approach ends up by relating quantum transport to representation theory of permutation and unitary groups, and how its success is related to the connection with areas of discrete mathematics and Selberg integrals, particularly through the use of Schur polynomials, for $\beta=2$. For other values of $\beta$, the role of these polynomials is assumed by a generalization known as the Jack polynomials. In the context of time delay statistics, this was explored in \cite{NovaesEPL}.

Joint distributions for transmission eigenvalues in the presence of non-ideal coupling such as that provided by tunnel barriers, were derived in \cite{Vidal,Jarosz2015} in terms of hypergeometric functions of matrix argument, for the case of a single barrier. This theory was then used in \cite{VivoMe} to derive finite-$M$ results for transport moments. Several works treated time delay in that situation, but the generalization of the Laguerre ensemble to this case appeared only in \cite{Grabsch1}.

\section{Semiclassical approach}

The semiclassical limit corresponds, formally, to letting $\hbar\to 0$. More physically, the number of wavelengths that fit inside the system becomes infinitely large. As discussed in the introduction, if this limit is considered at fixed openings, Ehrenfest time effects become important and random matrix theory cannot be expected to offer an effective description. Instead, to arrive at universality one must imagine  shrinking leads, in such a way that the semiclassical approximation is valid but still $N_1,N_2$ remain fixed.

Nevertheless, it is still possible to consider the channel numbers to be large and develop perturbative formulations for transport quantities that result in some infinite series in inverse powers of $M$. For example, the conductance for systems is intact time-reversal invariance is given by $\frac{N_1N_2}{M+1}=\frac{N_1N_2}{M}(1-\frac{1}{M}+\cdots)$. The first correction, in this case, is called weak localization. 

In the semiclassical regime, matrix elements are approximated in terms of classical quantities. For spectral statistics of closed systems, the Gutzwiller trace formula requires periodic orbits, which can be arbitrarily long and whose number grows exponentially with the period. In the present context, on the other hand, the required orbits are not periodic but scattering. Moreover, the problem is simplified by the presence of the dwell time $\tau_D$, which places an effective threshold of the length of relevant orbits (exceedingly long ones are suppressed).

Element $oi$ from the matrix $S$ is approximated by a sum over trajectories $\alpha$ that enter from the incoming channel $i$ and leave from the outgoing channel $o$, 
\be S_{oi}=\frac{1}{\sqrt{\tau_H}}\sum_{\alpha:i\to o}A_\alpha e^{i\mathcal{S}_\alpha/\hbar},\ee where $\mathcal{S}_\alpha(E)$ is the action of the trajectory and the prefactor $A_\alpha$ is related to its stability. The semiclassical treatment of conductance, for example, consists of writing 
\be{\rm Tr}(T)=\sum_{i,o}t_{oi}t^\dagger_{io}=\frac{1}{\tau_H}\sum_{i,o}\sum_{\alpha\sigma}A_\alpha e^{i\mathcal{S}_\alpha/\hbar}A^*_\sigma e^{-i\mathcal{S}_\sigma/\hbar},\ee where $\alpha$ and $\sigma$ are two trajectories that go from $i$ to $o$. This is a strongly oscillating function of the energy, and its local average can be calculated under a stationary phase approximation, which leads to the condition that $\alpha$ and $\sigma$ must be identical or at least extremely similar.

If these two trajectories are in fact identical, we have $\frac{1}{\tau_H}\sum_{i,o}\sum_{\alpha}|A_\alpha|^2$, and this sum can be evaluated according to the rule, established by Richter and Sieber \cite{Richter2002}, that summation over trajectories connecting fixed channels is equivalent to integration over the time, $\sum_{\alpha}|A_\alpha|^2=\int_0^\infty e^{-t/\tau_D}dt=\tau_D$. After summing over the incoming and outgoing channels, this gives the correct leading approximation to conductance, $\frac{N_1N_2}{M}$. 

Higher order corrections in $1/M$ come from pairs of orbits that are not exactly identical but are almost so, with action differences of order $\hbar$. The major breakthrough in the semiclassical approach was the realization that the mechanism behind such constructive interference between trajectories was the existence of self-crossings, also known as encounters, where a long trajectory runs nearly parallel or anti-parallel to itself for a short time. This implies the existence of a ``partner'' orbit that differs from the original one only in the small region of the encounter and possibly in its direction of motion along some arcs. \footnote{Refer to chapter Y in this volume}

The first correction to the average conductance, for example, comes from trajectories that have the following structure: $\alpha$ has a long arc, then goes around a loop, then another long arc (along such arcs the motion is actually chaotic); $\sigma$ is the same, except the loop is traversed in the opposite direction. Therefore, where one of them has a crossing, the other has an anti-crossing. These are now called Sieber-Richter pairs. These authors proceeded to integrate $|A_{\alpha}|^2e^{i\Delta S/\hbar}$ over phase space, and showed that the result factorizes nicely: arcs and loops result each in a factor $1/M$, and the encounter contributes a factor $(-M)$. This indeed results in the weak localization term, $-N_1N_2/M^2$.

This approach was then generalized by Haake and collaborators, who shortly produced the whole pertubative series for the average conductance \cite{Heusler2006}, showing that all corrections to the main result cancel out for systems without time-reversal symmetry, and for those with that symmetry the corrections produce precisely the geometric series that adds up to $N_1N_2/(M+1)$. Moreover, they generalized the notion of encounters to allow for multiple stretches of trajectories, and addressed other transport moments \cite{Muller2007}, like conductance variance and shot-noise \cite{HeuslerNoise}, using recursion arguments and summing infinite series to achieve agreement with random matrix theory. 

The semiclassical calculation of a transport moment like
\be\label{pn} {\rm Tr}(T^n)=\sum_{\vec{i}=1}^{N_1}\prod_{k=1}^n (t^\dagger t)_{i_{k+1},i_k}=\sum_{\vec{i}=1}^{N_1}\sum_{\vec{o}=1}^{N_2}\prod_{k=1}^n t^\dagger_{i_{k+1},o_k} t_{o_{k},i_k}\ee
requires multiple sums over trajectories,\be\label{multiple}
\frac{1}{\tau_H^n}\sum_{\vec{i},\vec{o}}\sum_{\alpha,\sigma}A_\alpha
A^*_\sigma e^{i(\mathcal{S}_\alpha - \mathcal{S}_\sigma)/\hbar},\ee with the understanding that $\alpha_k$ goes from $i_k$ to $o_k$, while $\sigma_k$ goes from $i_{k+1}$ to $o_k$. The quantity $A_\alpha=\prod_k A_{\alpha_k}$ is a collective stability, while $\mathcal{S}_\alpha=\sum_k
\mathcal{S}_{\alpha_k}$ it is a collective action of trajectories $\alpha$, and similarly for $\sigma$.

Under the stationary phase approximation, constructive interference requires that the set of trajectories $\alpha$  be extremely similar, as a group, to the set of trajectories $\sigma$. Again, it is encounters that play a central role. The result is naturally perturbative in $1/M$, and essentially diagrammatic: the contribution of a diagram with $L$ arcs/edges and $V$ encounters/vertices is proportional to $M^{V-L}$. The quantity $V-L$, in turn, is the Euler characteristic of the diagram/graph, indicating that the theory is topological in nature.

For example, we see in Figure 1 a diagram that contributes to the calculation of the average value of $({\rm Tr}(T))^2$. Two $\alpha$ trajectories are depicted in solid line, and two $\sigma$ ones in dashed line. There is one simple encounter and one triple encounter, both of them greatly magnified for clarity. The chaotic nature of the trajectories is not shown, also for clarity. Such a diagram is evaluated as $(-1)^2N_1^2N_2^2/M^{7-2}$, since it has $L=7$ and $V=2$.

\begin{figure}[t]
\centering
\includegraphics[width=.5\textwidth]{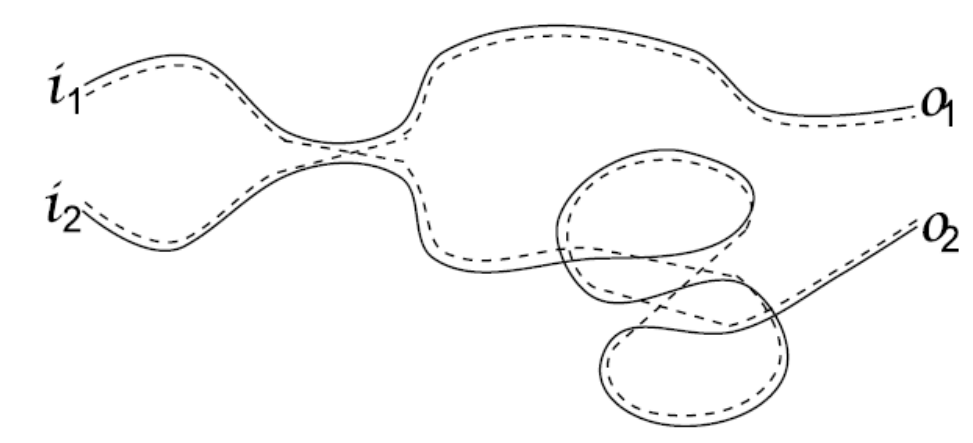}
\caption{Diagram that contributes to the average value of $({\rm Tr}(T))^2$, with one simple encounter and one triple encounter, both of them greatly magnified. Two $\alpha$ trajectories are depicted in solid line, and two $\sigma$ ones in dashed line. The chaotic nature of these trajectories is not shown. Taken from \cite{Novaes2015E}.}
\label{chap1:fig1}
\end{figure}

A leading-order calculation of the average value of (\ref{pn}) for all $n$ appeared in \cite{gregHarr}, and soon diagrammatical and group-theoretical arguments were used is such a way that equivalence between the semiclassical approach and random matrix theory was established for all transport moments \cite{greg,Berkolaiko2012}, under the universality regime we have been considering.

On the other hand, Ehrenfest-time effects were taken into account in calculations of weak localization \cite{Ehr2}, coherent backscattering \cite{EhrCoh}, shot-noise \cite{Ehr3}, counting statistics \cite{Ehr4} and even Andreev superconducting billiards \cite{Andreev}. Because the duration of the encounter is of the order of $\tau_E$, the diagrammatics has to change as now the vertices have internal structure. This leads to a theory which is not topological, i.e. although the trajectories are similar, the $\tau_E$-dependent contribution of a diagram is no longer determined just by its Euler characteristic, and this complicates matters enormously.

The versatility of the semiclassical approximation was further demonstrated when it was generalized to include tunnel barriers in the leads \cite{tun1,tun2,tun3}, a slightly more realistic situation in which the coupling between the internal part of the chaotic cavity and the outside is not perfectly transparent, corresponding to a finite reflection probability when the electron attempts to escape. Assuming that lead $a$ has reflection probability $R_a$, the diagrammatical contribution of each arc becomes $(M-R_1N_1-R_2N_2)^{-1}$, and the contribution of each vertex of valence $2q$ becomes $-M+R_1^qN_1+R_2^qN_2$. There is also a factor $R_a$ for each time a trajectory reflects off the barrier at lead $a$.

Another interesting generalization was the treatment of trajectories with different energies \cite{en1,en2}, allowing the calculation of some large$-M$ time-delay statistics, and the derivation of some correlations between $S$ matrices at different energies. For example, when considering something like 
\be\label{C} C_n(M,\epsilon)=\frac{1}{M}{\rm Tr}\left[S(E+\frac{\epsilon\hbar}{2\tau_D})S^\dagger(E-\frac{\epsilon\hbar}{2\tau_D})\right]^n,\ee the diagrammatical contribution of each arc becomes $[M(1-i\epsilon)]^{-1}$, and the contribution of each vertex of valence $2q$ becomes $-M(1-iq\epsilon)$. As a by product, the average value of time delay moments can then be obtained by taking derivatives, 
\be \frac{1}{M}{\rm Tr}(Q^m)=\frac{\tau_D^m}{i^mm!}\frac{d^m}{d\epsilon^m}\sum_{n=1}^m(-1)^{m-n}\binom{m}{n}C_n,\ee as discussed in \cite{en2,beyond}. 

However, a diagrammatic formalism dedicated specifically to time delay was developed in \cite{Kuipers2014}, and it circumvents the need for different energies. This more direct approach requires trajectories that enter the chaotic region but never leave, instead they terminate at some point inside the cavity. Perhaps unsurprisingly, the diagrammatic rules for computing the average value of traces of powers of the time delay matrix is that each arc contributes $1/M$ and the contribution of each vertex is $-M$ if it does not contain an end point. The difference with respect to the transport case is that now the vertex contributes $1$ if it contains one end point and $0$ if it contains more than one end point.

For example, we see in Figure 2 a diagram that contributes to the calculation of the average value of ${\rm Tr}(Q)$, with a single encounter. Notice how the trajectories enter the chaotic region but do not leave. Again, their chaotic nature is not shown. Such a diagram is evaluated as $-1/M^{3-1-1}$, since it involves one channel, three arcs and one encounter that does not contain an end point.

\begin{figure}[t]
\centering
\includegraphics[width=.5\textwidth]{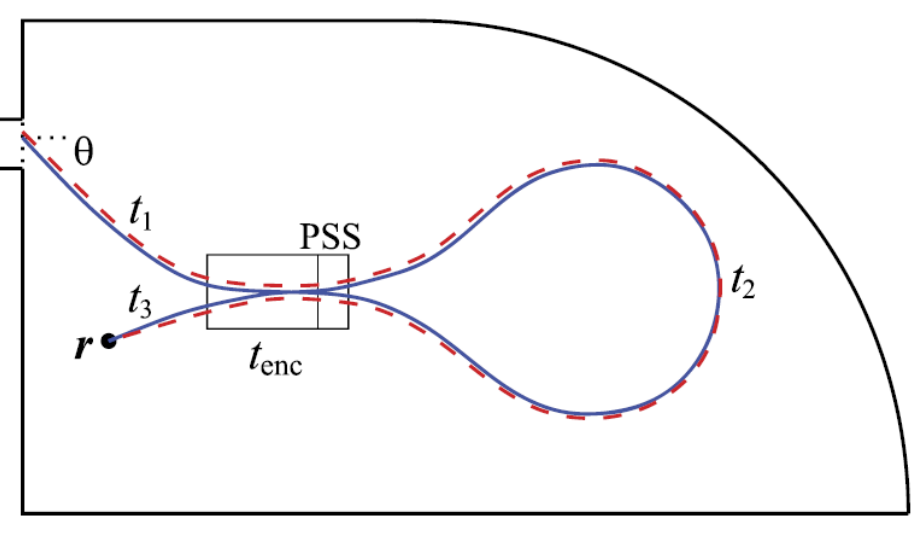}
\caption{Diagram that contributes to the average value of ${\rm Tr}(Q)$, with one simple encounter.  Since it involves one channel, three arcs and one encounter that does not contain an end point, its contribution is $-1/M$. Taken from \cite{Kuipers2014}.}
\label{chap1:fig2}
\end{figure}

\section{Matrix models for semiclassical transport}

\subsection{Transport}

As we have seen, the semiclassical approach to transport leads to a diagrammatic theory which is perturbative in $1/M$, as the contribution of a diagram is proportional to $1/M^{L-V}$, where $L$ is the number of edges and $V$ the number of vertices. As discussed in \cite{Novaes2013,Novaes2015b}, an auxiliary matrix integral can be defined that leads to precisely the same diagrammatic rules, providing a kind of integral representation for the transport moments (notice that these matrix integrals are not identical to the random matrix approach: the matrix being integrated over is a dummy variable and is not supposed to represent any physical quantity).

Suppose the matrix integral \be \langle f\rangle=\int dZ
e^{-M\tr (ZZ^\dag)}f(Z,Z^\dag),\ee where $M$ is a
parameter and $dZ$ denotes the flat measure on the space of $N$-dimensional matrices,
such that each matrix element is independently integrated over the whole complex plane.
Because of the Gaussian weight, we
 have \be\label{cov} \langle
Z_{mj}Z^\dag_{qr}\rangle=\frac{\delta_{mr}\delta_{jq}}{M}\ee and the so-called Wick's rule,
\begin{align}\label{wick1} \left\langle\prod_{k=1}^n
Z_{m_kj_k}Z^\dag_{q_{k}r_k}\right\rangle=\sum_{\sigma\in S_n}\prod_{k=1}^n\langle
Z_{m_kj_k}Z^\dag_{q_{\sigma(k)}r_{\sigma(k)}}\rangle=\frac{1}{M^n}\sum_{\sigma\in S_n}
\prod_{k=1}^n\delta_{m_kr_{\sigma(k)}}\delta_{j_{k}q_{\sigma(k)}},\end{align} is a
well known result \cite{zvon}. In words, it says we must sum, over all possible pairings
between $Z$'s and $Z^\dag$'s, the product of the average values of the pairs.

This leads to a diagrammatical approach \cite{diag1,diag2,diag3} in which $\tr(ZZ^\dag)^q$ is represented by a vertex of valence $2q$, and all possible connections among vertices are summed over, leading to all possible such diagrams. Each vertex is accompanied by a factor $(-M)$ and each edge by a factor $1/M$, exactly like in the semiclassical case. Thereby, the calculation of 
\be \left\langle \prod_{k=1}^n t_{i_ko_k}t^\dag_{o_{k+1}i_k}\right\rangle,\ee can be turned into something proportional to \be\label{model} \int dZ
e^{-M\sum_{q\ge 1} \frac{1}{q}\tr[(ZZ^\dag)^q]} \prod_{k=1}^n
Z_{i_ko_k}Z^\dag_{o_{k+1}i_k}.\ee  

However, when summing over all possible connections, closed loops may end up being produced, which would correspond to periodic orbits in the semiclassical calculation. Such orbits are not allowed. Fortunately, each closed loop is associate with a free index in the matrix elements \cite{Novaes2013,Novaes2015b}, and this leads to a factor $N$, so the relevant diagrams are those whose contribution is independent of $N$. Equivalently, the limit $N\to 0$ may be taken. 

This approach is very efficient, because the sum over all possible diagrams is performed automatically by the integral, which is computed using traditional techniques from matrix integrals like singular value decomposition \cite{mehta, mathai} and Weingarten functions \cite{collins,CS}. As a consequence, less cumbersome proofs of the semiclassical/RMT equivalence can be obtained \cite{Novaes2013,Novaes2015b} and the underlying algebraic properties of this equivalence are revealed. 

Moreover, these semiclassical matrix models have been generalized in different ways. 

Introducing a parameter $\epsilon$ as 
\be\label{withe} \int dZ
e^{-M\sum_{q\ge 1} \frac{1-iq\epsilon}{q}\tr[(ZZ^\dag)^q]} \prod_{k=1}^n
Z_{i_ko_k}Z^\dag_{o_{k+1}i_k}\ee 
leads to a diagrammatical formulation that is equivalent \cite{Novaes2015E} to the one developed \cite{en1,en2} for the semiclassical calculation of 
\be \left\langle \prod_{k=1}^n t_{i_ko_k}(E+\epsilon\hbar/2\tau_D)t^\dag_{o_{k+1}i_k}(E-\epsilon\hbar/2\tau_D)\right\rangle.\ee This can be used to derive results \cite{ED2,PRE105} that are either power series in $\epsilon$ with coefficients that are rational in $M$, or power series in $1/M$ with coefficients that are rational in $\epsilon$. 

On the other hand, introducing the parameter $R$ as
\be \int dZ
e^{-\sum_{q\ge 1} \frac{M-N_1R^q}{q}\tr[(ZZ^\dag)^q]} \prod_{k=1}^n
\left(Z\frac{1}{1-RZ^\dagger Z}\right)_{i_ko_k}Z^\dag_{o_{k+1}i_k}\ee 
leads to a diagrammatical formulation that is equivalent \cite{bento} to the one developed for the case when there is a tunnel barrier of reflection probability $R$ in the first lead. Results obtained in this way are equivalent to RMT ones, but are expressed using much simpler formulae. Currently, as for RMT, it is not possible to treat in a systematic way the presence of tunnel barriers in both leads, except for the simplest transport quantity, the average conductance \cite{conduc}, or in the case when both barriers are identical \cite{ident}.

\subsection{Time delay}

The matrix model (\ref{withe}) can be used to treat the correlations $C_n(M,\epsilon)$, see Eq.(\ref{C}), and these can then be converted into results for time delay moments. On the other hand, the adapted diagrammatic rules from \cite{Kuipers2014} can be encoded directly in a suitable matrix model \cite{Novaes2023PhysicaD}, namely something proportional to
\be \sum_{i_1,...,i_n}\int dZ
e^{-M\sum_{q\ge 1} \frac{1}{q}\tr[(ZZ^\dag)^q]} \prod_{k=1}^n
\left(Z\frac{1}{1-Z^\dagger Z}\right)_{i_k,k}Z^\dag_{k+1,i_k}.\ee
Taylor expansion of the exponential (except for the Gaussian part), produces all possible vertices that do not contain end points. The term $Z^\dag_{k+1,i_k}$ represents a trajectory going from incoming channel $i_k$ (which are summed over) to end point $r_k$, the term $\left(Z\frac{1}{1-Z^\dagger Z}\right)_{i_k,k}$ represents a trajectory going from incoming channel $i_{k+1}$ to end point $r_k$, and the geometric series produces vertices of any valence that do contain an end point.

Solving the above matrix integral by using the usual machinery (singular value decomposition, Schur polynomials, Weingarten functions) one can obtain semiclassical approximations to $\frac{1}{M}{\rm Tr}(Q^m)$ to high order in $1/M$. A complete proof of equivalence of RMT is still contingent on the demonstration of a particular combinatorial algebraic identity that is currently beyond reach.

Also in this setting it is interesting to take into account an imperfect coupling between
the scattering region and the exterior, such as a tunnel barrier of reflection probability $R$. 
This has no effect on the average time delay, and this can be understood semiclassically as follows: in
the presence of the barrier, an incident particle may be reflected promptly without delay, with probability $R$, or enter the cavity with probability $1-R$; after a time $\tau_D$, it tries to leave and succeeds with probability $1-R$ or is reflected back inside with probability $R$; and so on. Summing over all possibilities leads to a total average time delay which is
\be (1-R)(\tau_D+2\tau_D R+3\tau_D R^2+\cdots)(1-R)=\tau_D.\ee

The presence of the tunnel barrier of course influences higher statistics, as known from RMT results, so that in general $\langle s_\lambda(Q)\rangle(R)\neq \langle s_\lambda(Q)\rangle(0)$. For example, the average of the Wigner time delay squared is given by \cite{fyodorov,prl76}
\be\frac{\langle \tau_W^2\rangle}{\tau_D^2}=1+\frac{2(1-R^{M+1})}{(1-R)^2(M^2-1)}.\ee
This problem was taken up semiclassically in \cite{Novaes2023PRE}, where the semiclassical diagrammatic rules were derived and implemented in the form of a matrix model proportional to
\be \sum_{i_1,...,i_n}\int dZ
e^{-M\sum_{q\ge 1} \frac{1-R^q}{q}\tr[(ZZ^\dag)^q]} \prod_{k=1}^n
\left(Z\frac{1}{1-Z^\dagger Z}\right)_{i_k,k}\left(\frac{1}{1-Z^\dagger Z}Z^\dagger\right)_{k+1,i_k}.\ee
The solution of the above integral is more convoluted than previous simpler versions, but still explicit enough to be used in a computer to find the first few terms in power series in $R$ and $1/M$. 

For $\langle \tau_W^2\rangle/\tau_D^2$ the calculations are consistent with the result $1+\frac{2}{(1-R)^2(M^2-1)}$, missing the term containing $R^{M+1}$. This term is exponentially small for large $M$, so it is not surprising that a perturbative approach in $1/M$ fails to account for it. On the other hand, a very similar term appears in the calculation of average conductance, and it was obtained semiclassically in \cite{conduc} by means of a suitable regularization procedure. It is probable that something analogous could be done for time delay, but this is still a challenge.

The semiclassical approach in \cite{Novaes2023PRE} lead to some interesting conjectures about Schur-moments of time delay in the presence of a tunnel barrier, so let us finish by mentioning these open problems. Given an integer partition $\lambda$, let $\lambda'$ denote its conjugate (obtained by transposing the Young diagram of $\lambda$). Then it seems that moments corresponding to self-conjugate partitions are independent of $R$, so 
\be \langle s_\lambda(Q)\rangle(R)= \langle s_\lambda(Q)\rangle(0), \text{ if }\lambda=\lambda'.\ee Also, two reciprocity relations seem to hold:
\be \langle s_{\lambda'}(Q)\rangle(R,-M)=(-1)^{|\lambda|}\langle s_\lambda(Q)\rangle(R,M)\ee
and
\be [M]^\lambda\langle s_{\lambda'}(Q)\rangle(R^{-1},M)=[M]_\lambda\langle s_\lambda(Q)\rangle(R,M).\ee
None of these relations has been proved, either with RMT or semiclassics. Their simplicity indicates that the average Schur polynomials are indeed the most natural moments associated with a random matrix when $\beta=2$ (their role is probably taken up by Jack polynomials for general $\beta$; whether analogous relations hold is an interesting question).

\section{Conclusion}

The semiclassical approach to the quantum properties of chaotic systems underwent intense development in the 21st century, after the realization that encounters are the fundamental mechanism behind the existence of correlated trajectories that lead to systematic constructive interference \cite{scripta}. Taking these structures into account rapidly led to the development of perturbative treatments that exhibited perfect agreement with random matrix theory and a much more thorough understanding of the equivalence of these two approaches (when such equivalence exists). Not only that, but the theory proved to be versatile enough to eventually accommodate a variety of extra elements, like Ehrenfest time dependence, trajectories with different energies, tunnel barriers, Andreev reflections. 

Because encounters can be described by permutations \cite{perms}, a whole algebraic machinery can be brought to bear upon the problem, alongside more visual diagrammatical methods. This allows, for example, connections to be made with factorization enumeration, recursion relations and representation theory.

A further boost came with the introduction of the matrix integral representations discussed in the previous section, not only for transport moments and time delay moments, but also spectral correlations \cite{eumuller}. On the one hand, this approach makes it much more natural to prove equivalence with random matrix theory, since the calculation is already written as a matrix integral in the first place. On the other hand, powerful techniques from matrix integration can be employed in order to arrive at explicit solutions, and in some cases even go beyond random matrix theory.

\bibliographystyle{JHEP}%
\bibliography{SemicTranspRefs}

\end{document}